\pdfoutput=1

\documentclass[12pt,a4paper]{article}

\usepackage{ifthen} 
\newboolean{pdflatex}
\setboolean{pdflatex}{true} 

\newboolean{articletitles}
\setboolean{articletitles}{true} 

\newboolean{uprightparticles}
\setboolean{uprightparticles}{false} 

\newboolean{inbibliography}
\setboolean{inbibliography}{false} 

\def\paperauthors{LHCb collaboration} 
\def\paperasciititle{Measurement of the ratio of branching fractions  of the decays Lb2psi(2S)L and Lb2J/psiL} 
\def\papertitle{Measurement of the ratio of branching fractions  of the decays \psichannel  and \jpsichannel} 
\def\paperkeywords{{High Energy Physics}, {LHCb}} 
\def\papercopyright{\the\year\ CERN for the benefit of the LHCb collaboration} 
\def\paperlicence{CC-BY-4.0 licence}
\def\paperlicenceurl{https://creativecommons.org/licenses/by/4.0/}

\usepackage[top=1in, bottom=1.25in, left=1in, right=1in]{geometry}

\usepackage{microtype}
\usepackage{lineno}  
\usepackage{xspace} 
\usepackage{caption} 

\usepackage{graphicx}  
\usepackage{color}
\usepackage{colortbl}
\graphicspath{{./figs/}} 

\usepackage{amsmath} 
\usepackage{amssymb}
\usepackage{amsfonts}
\usepackage{upgreek} 

\newcommand*\patchAmsMathEnvironmentForLineno[1]{%
\expandafter\let\csname old#1\expandafter\endcsname\csname #1\endcsname
\expandafter\let\csname oldend#1\expandafter\endcsname\csname
end#1\endcsname
 \renewenvironment{#1}%
   {\linenomath\csname old#1\endcsname}%
   {\csname oldend#1\endcsname\endlinenomath}%
}
\newcommand*\patchBothAmsMathEnvironmentsForLineno[1]{%
  \patchAmsMathEnvironmentForLineno{#1}%
  \patchAmsMathEnvironmentForLineno{#1*}%
}
\AtBeginDocument{%
\patchBothAmsMathEnvironmentsForLineno{equation}%
\patchBothAmsMathEnvironmentsForLineno{align}%
\patchBothAmsMathEnvironmentsForLineno{flalign}%
\patchBothAmsMathEnvironmentsForLineno{alignat}%
\patchBothAmsMathEnvironmentsForLineno{gather}%
\patchBothAmsMathEnvironmentsForLineno{multline}%
\patchBothAmsMathEnvironmentsForLineno{eqnarray}%
}

\usepackage{hyperxmp}

\usepackage[pdftex,
            pdfauthor={\paperauthors},
            pdftitle={\paperasciititle},
            pdfkeywords={\paperkeywords},
            pdfcopyright={Copyright (C) \papercopyright},
            pdflicenseurl={\paperlicenceurl}]{hyperref}

\usepackage[all]{hypcap} 

\usepackage{xspace} 
\usepackage{upgreek}

\def\lhcb {\mbox{LHCb}\xspace}
\def\atlas  {\mbox{ATLAS}\xspace}

\def\velo   {VELO\xspace}

\def\MagUp {\mbox{\em Mag\kern -0.05em Up}\xspace}

\ifthenelse{\boolean{uprightparticles}}%
{

 \def\Pmu         {\ensuremath{\upmu}\xspace}

 \def\Ppi         {\ensuremath{\uppi}\xspace}

 \def\Ppsi        {\ensuremath{\uppsi}\xspace}

 \def\PDelta      {\ensuremath{\Delta}\xspace}                 
 \def\PXi      {\ensuremath{\Xi}\xspace}                 
 \def\PLambda      {\ensuremath{\Lambda}\xspace}                 
 \def\PSigma      {\ensuremath{\Sigma}\xspace}                 
 \def\POmega      {\ensuremath{\Omega}\xspace}                 
 \def\PUpsilon      {\ensuremath{\Upsilon}\xspace}                 
 
 \def\PB      {\ensuremath{\mathrm{B}}\xspace}                 
                  
 \def\PD      {\ensuremath{\mathrm{D}}\xspace}

 \def\PJ      {\ensuremath{\mathrm{J}}\xspace}                 
 \def\PK      {\ensuremath{\mathrm{K}}\xspace}

 \def\PX      {\ensuremath{\mathrm{X}}\xspace}

 \def\Pb      {\ensuremath{\mathrm{b}}\xspace}                 
 \def\Pc      {\ensuremath{\mathrm{c}}\xspace}                 
                  
 \def\Pe      {\ensuremath{\mathrm{e}}\xspace}

 \def\Pi      {\ensuremath{\mathrm{i}}\xspace}

 \def\Pp      {\ensuremath{\mathrm{p}}\xspace}

}
{

 \def\Pmu         {\ensuremath{\mu}\xspace}

 \def\Ppi         {\ensuremath{\pi}\xspace}

 \def\Ppsi        {\ensuremath{\psi}\xspace}                 
                  
 \mathchardef\PDelta="7101
 \mathchardef\PXi="7104
 \mathchardef\PLambda="7103
 \mathchardef\PSigma="7106
 \mathchardef\POmega="710A
 \mathchardef\PUpsilon="7107
                  
 \def\PB      {\ensuremath{B}\xspace}                 
                  
 \def\PD      {\ensuremath{D}\xspace}

 \def\PJ      {\ensuremath{J}\xspace}                 
 \def\PK      {\ensuremath{K}\xspace}

 \def\PX      {\ensuremath{X}\xspace}

 \def\Pb      {\ensuremath{b}\xspace}                 
 \def\Pc      {\ensuremath{c}\xspace}                 
                  
 \def\Pe      {\ensuremath{e}\xspace}

 \def\Pi      {\ensuremath{i}\xspace}

 \def\Pp      {\ensuremath{p}\xspace}

}

\makeatletter
\ifcase \@ptsize \relax
  \newcommand{\miniscule}{\@setfontsize\miniscule{4}{5}}
\or
  \newcommand{\miniscule}{\@setfontsize\miniscule{5}{6}}
\or
  \newcommand{\miniscule}{\@setfontsize\miniscule{5}{6}}
\fi
\makeatother

\DeclareRobustCommand{\optbar}[1]{\shortstack{{\miniscule (\rule[.5ex]{1.25em}{.18mm})}
  \\ [-.7ex] $#1$}}

\def\epem       {{\ensuremath{\Pe^+\Pe^-}}\xspace}

\def\mumu       {{\ensuremath{\Pmu^+\Pmu^-}}\xspace}

\def\lepton     {{\ensuremath{\ell}}\xspace}

\def\cquark    {{\ensuremath{\Pc}}\xspace}
\def\cquarkbar {{\ensuremath{\overline \cquark}}\xspace}

\def\bquark    {{\ensuremath{\Pb}}\xspace}

\def\pion   {{\ensuremath{\Ppi}}\xspace}

\def\pim    {{\ensuremath{\pion^-}}\xspace}

\def\kaon    {{\ensuremath{\PK}}\xspace}
  \def\Kbar    {{\kern 0.2em\overline{\kern -0.2em \PK}{}}\xspace}

\def\KorKbar    {\kern 0.18em\optbar{\kern -0.18em K}{}\xspace}

\def\Kp      {{\ensuremath{\kaon^+}}\xspace}

\def\KS      {{\ensuremath{\kaon^0_{\mathrm{ \scriptscriptstyle S}}}}\xspace}

  \def\Dbar    {{\kern 0.2em\overline{\kern -0.2em \PD}{}}\xspace}

\def\DorDbar    {\kern 0.18em\optbar{\kern -0.18em D}{}\xspace}

\def\B       {{\ensuremath{\PB}}\xspace}
\def\Bbar    {{\ensuremath{\kern 0.18em\overline{\kern -0.18em \PB}{}}}\xspace}

\def\BorBbar    {\kern 0.18em\optbar{\kern -0.18em B}{}\xspace}

\def\Bu      {{\ensuremath{\B^+}}\xspace}

\def\Bd      {{\ensuremath{\B^0}}\xspace}

\def\jpsi     {{\ensuremath{{\PJ\mskip -3mu/\mskip -2mu\Ppsi\mskip 2mu}}}\xspace}
\def\psitwos  {{\ensuremath{\Ppsi{(2S)}}}\xspace}

  \def\Y#1S{\ensuremath{\PUpsilon{(#1S)}}\xspace}

\def\proton      {{\ensuremath{\Pp}}\xspace}

\def\Xires       {{\ensuremath{\PXi}}\xspace}
\def\Xiresm      {{\ensuremath{\Xires^-}}\xspace}

\def\Lz          {{\ensuremath{\PLambda}}\xspace}
\def\Lbar        {{\ensuremath{\kern 0.1em\overline{\kern -0.1em\PLambda}}}\xspace}
\def\LorLbar    {\kern 0.18em\optbar{\kern -0.18em \PLambda}{}\xspace}
\def\Lambdares   {{\ensuremath{\PLambda}}\xspace}

\def\Lb      {{\ensuremath{\Lz^0_\bquark}}\xspace}

\def\Xibm    {{\ensuremath{\Xires^-_\bquark}}\xspace}

\newcommand{\decay}[2]{\ensuremath{#1\!\to #2}\xspace}         

\def\to                 {\ensuremath{\rightarrow}\xspace}

\def\AT#1     {\ensuremath{A_{\mathrm{T}}^{#1}}\xspace}           

\def\C#1      {\ensuremath{\mathcal{C}_{#1}}\xspace}                       
\def\Cp#1     {\ensuremath{\mathcal{C}_{#1}^{'}}\xspace}                    
\def\Ceff#1   {\ensuremath{\mathcal{C}_{#1}^{\mathrm{(eff)}}}\xspace}        
\def\Cpeff#1  {\ensuremath{\mathcal{C}_{#1}^{'\mathrm{(eff)}}}\xspace}       
\def\Ope#1    {\ensuremath{\mathcal{O}_{#1}}\xspace}                       
\def\Opep#1   {\ensuremath{\mathcal{O}_{#1}^{'}}\xspace}                    

\newcommand{\tev}{\ifthenelse{\boolean{inbibliography}}{\ensuremath{~T\kern -0.05em eV}}{\ensuremath{\mathrm{\,Te\kern -0.1em V}}}\xspace}
\newcommand{\gev}{\ensuremath{\mathrm{\,Ge\kern -0.1em V}}\xspace}
\newcommand{\mev}{\ensuremath{\mathrm{\,Me\kern -0.1em V}}\xspace}
\newcommand{\kev}{\ensuremath{\mathrm{\,ke\kern -0.1em V}}\xspace}
\newcommand{\ev}{\ensuremath{\mathrm{\,e\kern -0.1em V}}\xspace}
\newcommand{\gevc}{\ensuremath{{\mathrm{\,Ge\kern -0.1em V\!/}c}}\xspace}
\newcommand{\mevc}{\ensuremath{{\mathrm{\,Me\kern -0.1em V\!/}c}}\xspace}
\newcommand{\gevcc}{\ensuremath{{\mathrm{\,Ge\kern -0.1em V\!/}c^2}}\xspace}
\newcommand{\gevgevcccc}{\ensuremath{{\mathrm{\,Ge\kern -0.1em V^2\!/}c^4}}\xspace}
\newcommand{\mevcc}{\ensuremath{{\mathrm{\,Me\kern -0.1em V\!/}c^2}}\xspace}

\def\mum  {\ensuremath{{\,\upmu\mathrm{m}}}\xspace}

\def\ps   {\ensuremath{{\mathrm{ \,ps}}}\xspace}

\newcommand{\stat}{\ensuremath{\mathrm{\,(stat)}}\xspace}
\newcommand{\syst}{\ensuremath{\mathrm{\,(syst)}}\xspace}

\newcommand{\chisqndf}{\ensuremath{\chi^2/\mathrm{ndf}}\xspace}

\def\gsim{{~\raise.15em\hbox{$>$}\kern-.85em
          \lower.35em\hbox{$\sim$}~}\xspace}
\def\lsim{{~\raise.15em\hbox{$<$}\kern-.85em
          \lower.35em\hbox{$\sim$}~}\xspace}

\def\ptot       {\mbox{$p$}\xspace}
\def\pt         {\mbox{$p_{\mathrm{ T}}$}\xspace}

\def\evtgen     {\mbox{\textsc{EvtGen}}\xspace}

\def\geant      {\mbox{\textsc{Geant4}}\xspace}

\def\pythia     {\mbox{\textsc{Pythia}}\xspace}

\def\tell1  {TELL1\xspace}
\def\ukl1   {UKL1\xspace}

\usepackage{cite} 
\usepackage{mciteplus}

\renewcommand{\L}{\Lambdares}
\newcommand{\xichannel}{\decay{\Xibm} {\jpsi \Xiresm}}
\newcommand{\xipsichannel}{\decay{\Xibm} {\psitwos \Xiresm}}
\newcommand{\xiccchannel}{\decay{\Xibm} {\Ppsi \Xiresm}}

\newcommand{\psichannel}{\decay{\Lb} {\psitwos \Lambdares}}
\newcommand{\jpsichannel}{\decay{\Lb}{\jpsi \Lambdares}}
\newcommand{\bdchannel}{\decay{\Bd}{\jpsi \KS}}
\newcommand{\bdccchannel}{\decay{\Bd}{\Ppsi \KS}}
\newcommand{\bdpsichannel}{\decay{\Bd}{\psitwos \KS}}
\newcommand{\buchannel}{\decay{\Bu}{\jpsi \Kp}}
\newcommand{\bupsichannel}{\decay{\Bu}{\psitwos \Kp}}

\usepackage{booktabs}
\usepackage[capitalise]{cleveref} 
\crefname{section}{Sec.}{Secs.}
\crefname{figure}{Fig.}{Figs.}
\crefname{chapter}{Chap.}{Chaps.}
\crefname{equation}{Eq.}{Eqs.}

\begin{document}

\renewcommand{\thefootnote}{\fnsymbol{footnote}}
\setcounter{footnote}{1}


\begin{titlepage}
\pagenumbering{roman}

\vspace*{-1.5cm}
\centerline{\large EUROPEAN ORGANIZATION FOR NUCLEAR RESEARCH (CERN)}
\vspace*{1.5cm}
\noindent
\begin{tabular*}{\linewidth}{lc@{\extracolsep{\fill}}r@{\extracolsep{0pt}}}
\ifthenelse{\boolean{pdflatex}}
{\vspace*{-2.7cm}\mbox{\!\!\!\includegraphics[width=.14\textwidth]{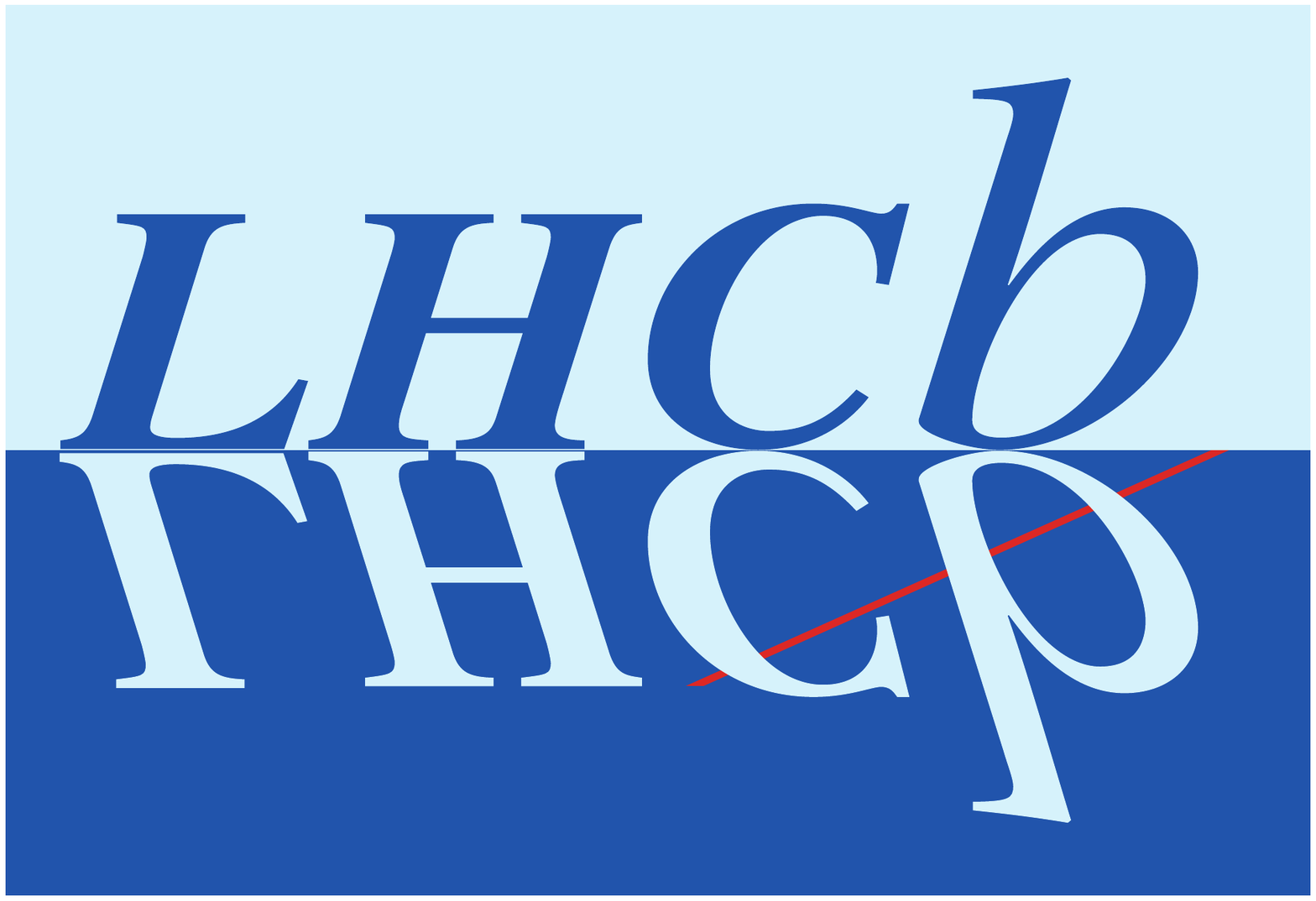}} & &}%
{\vspace*{-1.2cm}\mbox{\!\!\!\includegraphics[width=.12\textwidth]{lhcb-logo.eps}} & &}%
\\
 & & CERN-EP-2019-003 \\  
 & & LHCb-PAPER-2017-041 \\  
 & & 22 March 2019 \\ 
 & & \\
\end{tabular*}

\vspace*{4.0cm}

{\normalfont\bfseries\boldmath\huge
\begin{center}
  \papertitle 
\end{center}
}

\vspace*{2.0cm}

\begin{center}
\paperauthors\footnote{Authors are listed at the end of this paper.}
\end{center}

\vspace{\fill}

\begin{abstract}
  \noindent
Using  $pp$ collisions corresponding to 3$\,\text{fb}^{-1}$ integrated luminosity, recorded by the LHCb experiment at centre-of-mass energies of 7 and 8$\,\text{TeV}$,
the ratio of branching fractions
\begin{align*}
\mathcal{B}(\Lambda_b^0 \!\to \psi(2S) \Lambda)/\mathcal{B}(\Lambda_b^0 \!\to J\!/\!\psi \Lambda)=  0.513 \pm 0.023 \,(\text{stat}) \pm 0.016 \,(\text{syst}) \pm 0.011\, (\mathcal{B})
\end{align*}
is determined. The first uncertainty is statistical, the second is systematic and the third is due to the external branching fractions used.
  
\end{abstract}

\vspace*{2.0cm}

\begin{center}
Published in JHEP 03 (2019) 126
\end{center}

\vspace{\fill}

{\footnotesize 
\centerline{\copyright~\papercopyright, licence \href{\paperlicenceurl}{\paperlicence}.}}
\vspace*{2mm}

\end{titlepage}


\newpage
\setcounter{page}{2}
\mbox{~}
%
%
%
%

\cleardoublepage


\renewcommand{\thefootnote}{\arabic{footnote}}
\setcounter{footnote}{0}



\pagestyle{plain} 
\setcounter{page}{1}
\pagenumbering{arabic}


%

\section{Introduction}
The LHCb collaboration has observed many \decay{\Lb}{\jpsi \PX} \cite{LHCb-PAPER-2012-057,LHCb-PAPER-2014-020,LHCb-PAPER-2015-029,LHCb-PAPER-2015-032,LHCb-PAPER-2015-060,LHCb-PAPER-2016-009,LHCb-PAPER-2016-015} and \decay{\Lb}{\psitwos \PX} decays \cite{LHCb-PAPER-2015-060,LHCb-PAPER-2018-022}, where \PX indicates a final-state particle system.
Ratios of branching fractions of $\bquark$-hadron decays into $\psitwos \PX$ and $\jpsi \PX$ provide useful information on the production of charmonia in $\bquark$-hadron decays.
These ratios can be used to test factorisation of amplitudes.
The \atlas collaboration has previously measured the ratio of the branching fractions to be \mbox{$\mathcal{B}(\psichannel )/\mathcal{B}(\jpsichannel )=0.501 \pm 0.033\stat \pm 0.019 \syst$ \cite{Aad:2015msa}}.
This result differs by 2.8$\,\sigma$ from a theoretical prediction in the framework of the covariant quark model, $\mathcal{B}(\psichannel )/\mathcal{B}(\jpsichannel )=0.8\pm0.1$\cite{Gutsche:2013oea,Gutsche:2015lea}.
Variations of the used form factors \cite{PhysRevD.80.094016,doi:10.1142/S0217751X12500169} lead to predictions in the range of 0.65 to 1.14~\cite{Gutsche:2015lea}.
Also the result differs significantly from similar measurements in the \PB systems, $\mathcal{B}(\bdpsichannel )/\mathcal{B}(\bdchannel )=0.66\pm0.06$ and $\mathcal{B}(\bupsichannel )/\mathcal{B}(\buchannel )=0.615\pm0.019$\cite{PDG2018}.

In this paper the measurement of the branching fraction of the decay $\psichannel$ by \lhcb is presented.
Throughout this paper, the notation of a decay always implies the inclusion of the charge-conjugate process.
Determining the branching fraction of $\psichannel$ decays relative to the branching fraction of $\jpsichannel$ cancels most experimental uncertainties.
A measurement with improved precision helps to better understand this possible discrepancy and sets new constraints on the available form-factor models \cite{Gutsche:2015lea}.

\section{\lhcb detector}
The \lhcb detector~\cite{Alves:2008zz,LHCb-DP-2014-002} is a single-arm forward
spectrometer covering the \mbox{pseudorapidity} range $2<\eta <5$,
designed for the study of particles containing \bquark or \cquark
quarks. The detector includes a high-precision tracking system
consisting of a silicon-strip vertex detector (VELO) surrounding the $pp$
interaction region~\cite{LHCb-DP-2014-001}, a large-area silicon-strip detector located
upstream of a dipole magnet with a bending power of about
$4{\mathrm{\,Tm}}$, and three stations of silicon-strip detectors and straw
drift tubes~\cite{LHCb-DP-2013-003} placed downstream of the magnet.
The tracking system provides a measurement of the momentum, \ptot, of charged particles with
a relative uncertainty that varies from 0.5\% at low momentum to 1.0\% at 200\gevc.
The minimum distance of a track to a primary vertex (PV), the impact parameter (IP), 
is measured with a resolution of $(15+29/\pt)\mum$,
where \pt is the component of the momentum transverse to the beam, in\,\gevc.
Different types of charged hadrons are distinguished using information
from two ring-imaging Cherenkov detectors~\cite{LHCb-DP-2012-003}. 
Photons, electrons and hadrons are identified by a calorimeter system consisting of
scintillating-pad and preshower detectors, an electromagnetic
calorimeter and a hadronic calorimeter. Muons are identified by a
system composed of alternating layers of iron and multiwire
proportional chambers~\cite{LHCb-DP-2012-002}.
The online event selection is performed by a trigger~\cite{LHCb-DP-2012-004}, 
which consists of a hardware stage, based on information from the muon
system, followed by a software stage, which applies a full event
reconstruction.
In the simulation, $pp$ collisions are generated using \pythia~\cite{Sjostrand:2007gs,*Sjostrand:2006za} with a specific \lhcb configuration~\cite{LHCb-PROC-2010-056}.
Decays of hadronic particles are described by \evtgen~\cite{Lange:2001uf}.
The interaction of the generated particles with the detector, and its response, are implemented using the \geant toolkit~\cite{Allison:2006ve, *Agostinelli:2002hh} as described in Ref.~\cite{LHCb-PROC-2011-006}.

\section{Event selection and selection efficiencies}
The \jpsi and \psitwos charmonium states, collectively called \Ppsi, are reconstructed through their decay into two muons.
Two tracks not originating from any PV, that are identified as oppositely charged muons, are required to form a good vertex.
These muons have to fulfil various trigger requirements.
At the hardware stage an event is required to contain a muon with high \pt or two muons with a large product of their respective \pt values.
At the software stage further requirements are placed on the \pt , momenta and IP of the muons.
The reconstructed \Ppsi masses must be within $\pm100\mevcc$ of their known masses \cite{PDG2018}.

The \L candidates are reconstructed by combining a pion and a proton candidate.
Due to its long lifetime, the \L baryon can decay either inside or outside the VELO.
The pion and proton can be reconstructed including hits from the \velo (long track) or without (downstream track).
Combinations where the track types of pion and proton differ are not considered.
Due to different momentum resolutions of these track types, some selection requirements differ between the two samples.
The pion and proton candidates are required to have high momentum ($>2\gevc$) and high \pt and the tracks must be displaced from any PV.
In addition, long-track proton candidates must be consistent with the proton hypothesis.
The invariant mass of the pion and proton combination has to be compatible with the known \L mass \cite{PDG2018} and both tracks must come from a common vertex.
Furthermore, the \L candidate is required to have a decay time longer than $2\ps$.

The \Lb candidate is reconstructed by combining the \Ppsi and the \L candidates and requiring that they form a common vertex.
The PV that fits best to the \Lb flight direction is assigned as associated PV.
It is required that the \Lb momentum points back to this PV and its decay vertex is significantly displaced from this PV.
Additional requirements are imposed using a kinematic fit with constrained \Ppsi and \L masses.
For downstream-track candidates the reconstructed \L decay time using this fit must be longer than $9\ps$.
The \chisqndf of this kinematic fit is required to be smaller than $36/6$ for long-track candidates and smaller than $26/6$ for downstream-track candidates.

After the selection, about 1\% of all events contain multiple candidates.
Among these multiple candidates a single candidate is retained using a random but reproducible procedure.
To ensure a precise efficiency determination, fiducial cuts on the \Lb baryon, $\pt(\Lb)<20\gevc$ and $2<\eta(\Lb)<4.5$ are applied.

\begin{figure}
\includegraphics[width=.5\textwidth]{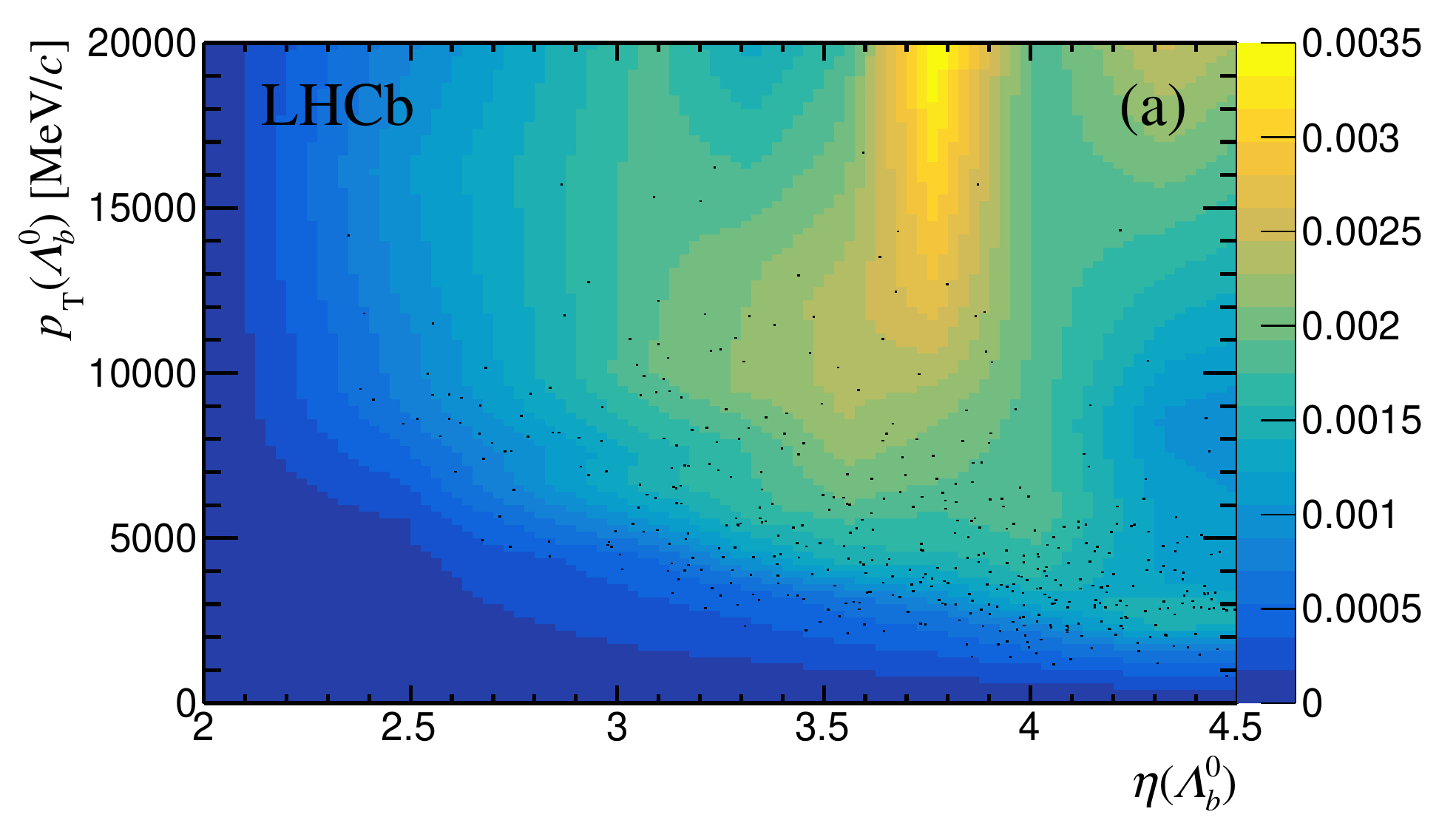}
\includegraphics[width=.5\textwidth]{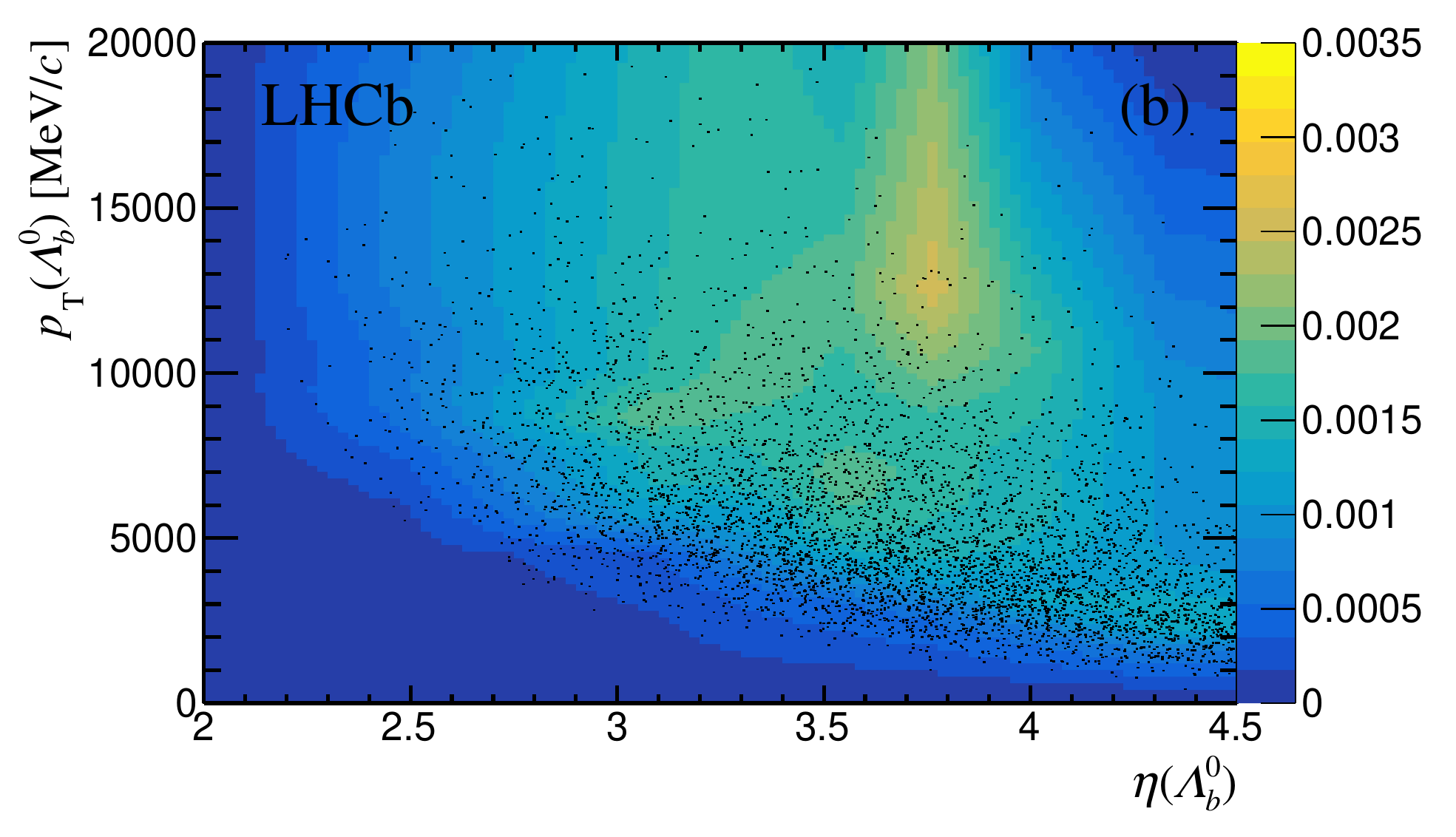}
\includegraphics[width=.5\textwidth]{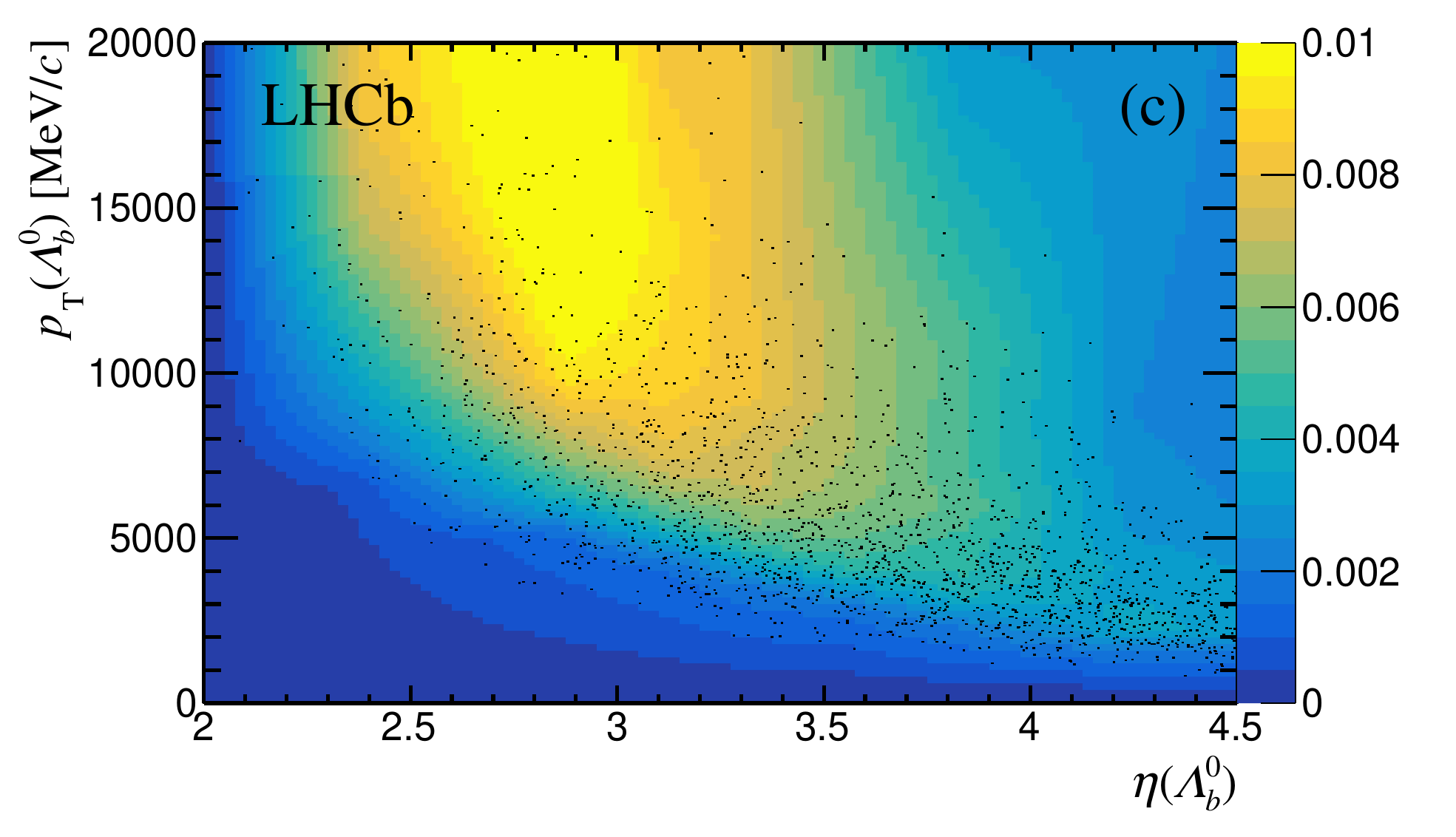}
\includegraphics[width=.5\textwidth]{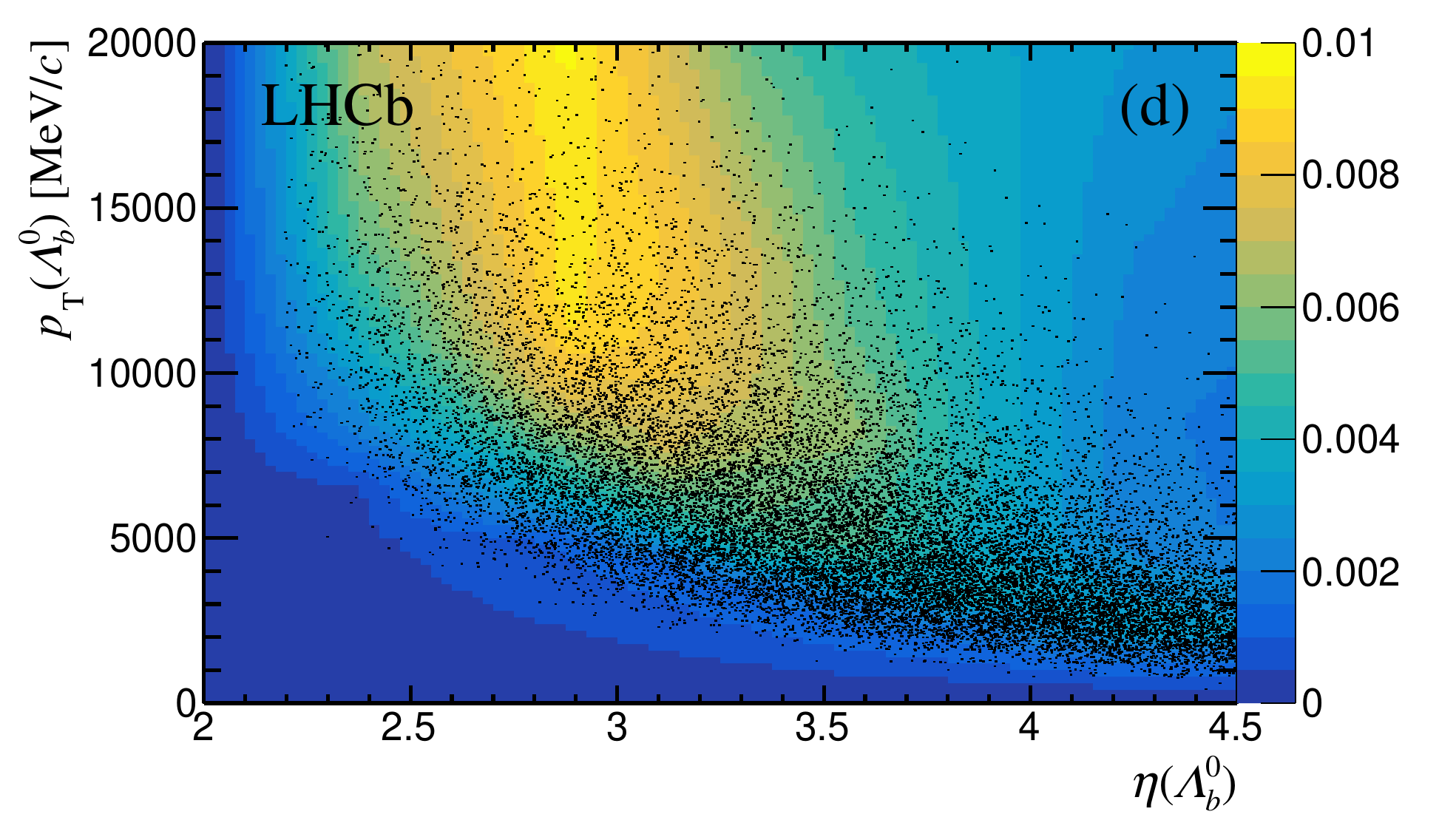}
\caption{Interpolated efficiency function for long-track candidates for (a) \psichannel and (b) \jpsichannel and for downstream-track candidates for (c) \psichannel and (d) \jpsichannel candidates. The distribution of the candidates on data is shown with black dots (each dot refers to one candidate).
Statistical fluctuations of the simulated sample are contributing to the efficiency function at the phase-space boundaries, where data candidates are not affected.}
\label{plot_efficiencies}
\end{figure}

The signal efficiency is evaluated separately for each channel and track type, using simulations and crosschecked with data.
The simulation assumes unpolarised decays but is corrected using theory predictions \cite{Gutsche:2013oea} for both decay channels.
Sources of inefficiencies are the geometrical acceptance of the detector, the trigger, the track reconstruction, and the candidate selection.
The last three efficiencies depend on the kinematics of the \Lb baryon, which is not perfectly simulated.
To account for the mismodelling, these efficiencies are determined in bins of $\pt(\Lb)$ and $\eta(\Lb)$.
The same binning scheme, consisting of seven bins for each of the two variables, is used for both decay channels.
The binning scheme is designed such that all bins are uniformly populated, with at least 100 entries in each bin.
The resulting efficiency for a given candidate is determined by linear interpolation of the binned efficiency model to reduce effects arising from the choice of the binning scheme.
For the interpolation, the mean value in each bin is used and additional bins are added to ensure interpolation at the boundaries.
The resulting efficiency functions together with the distribution of the corresponding signal candidates are shown in \cref{plot_efficiencies}.

\section{Signal yield determination}
\begin{figure}
\includegraphics[width=.5\textwidth]{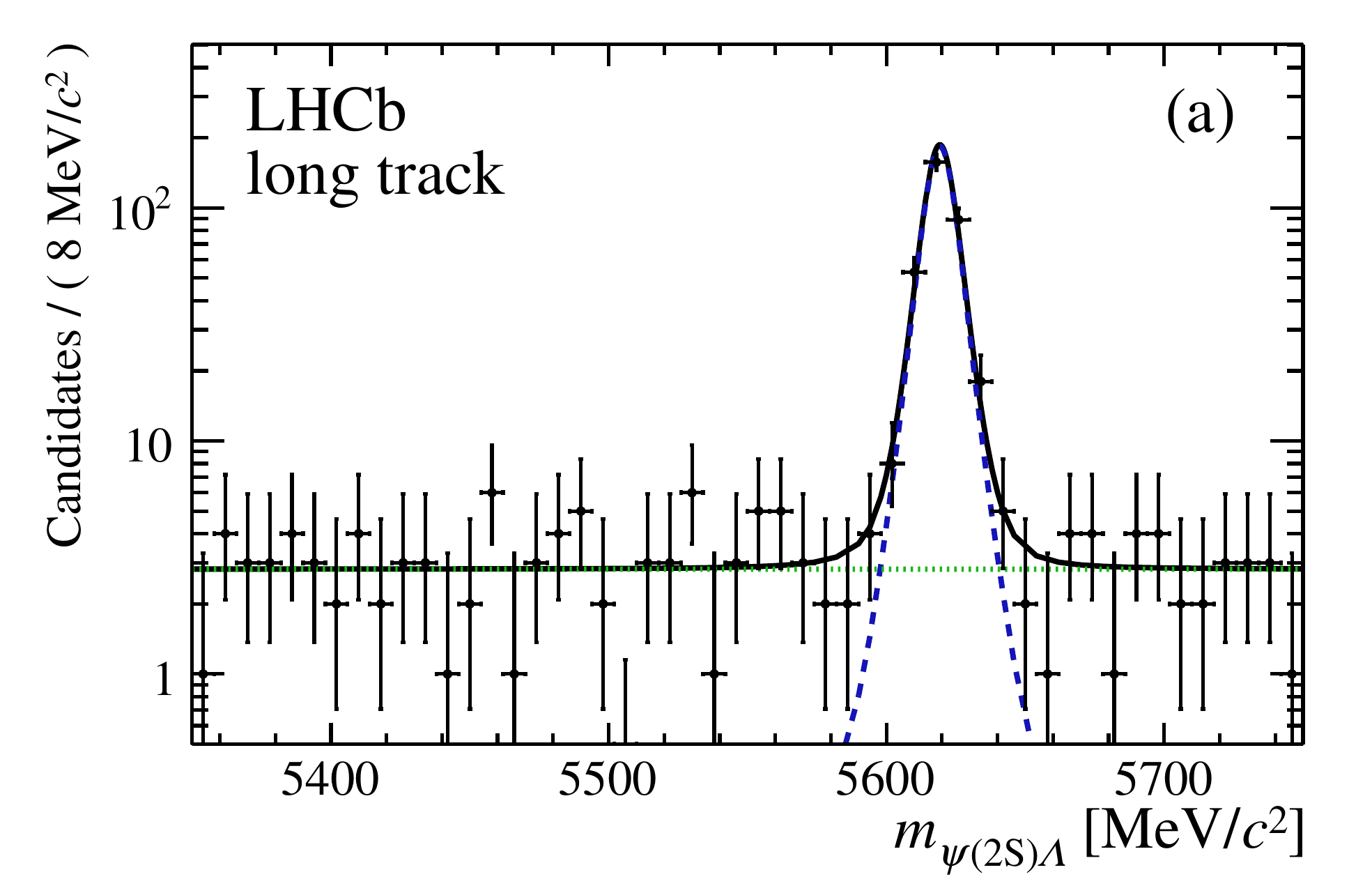}
\includegraphics[width=.5\textwidth]{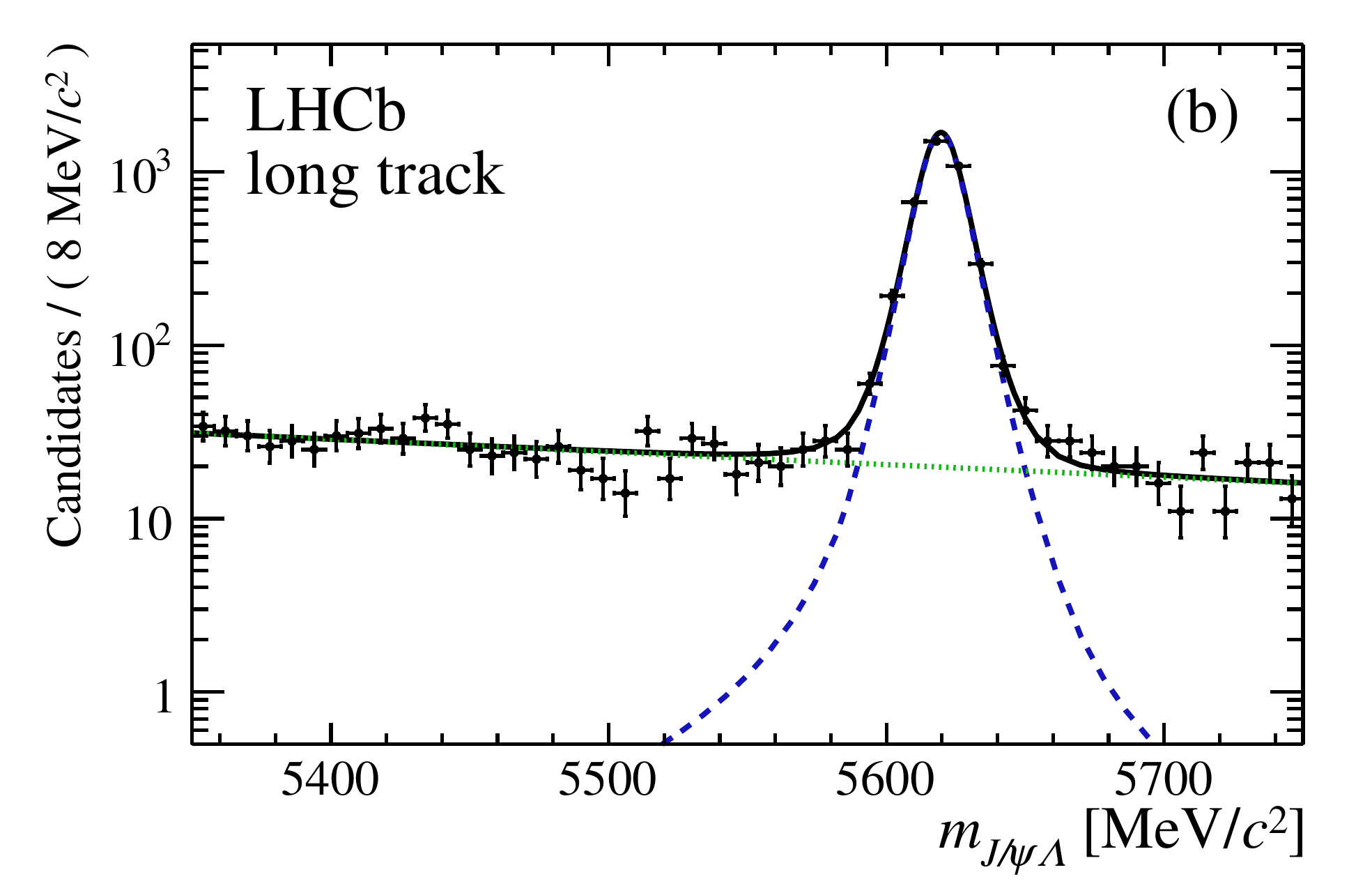}
\includegraphics[width=.5\textwidth]{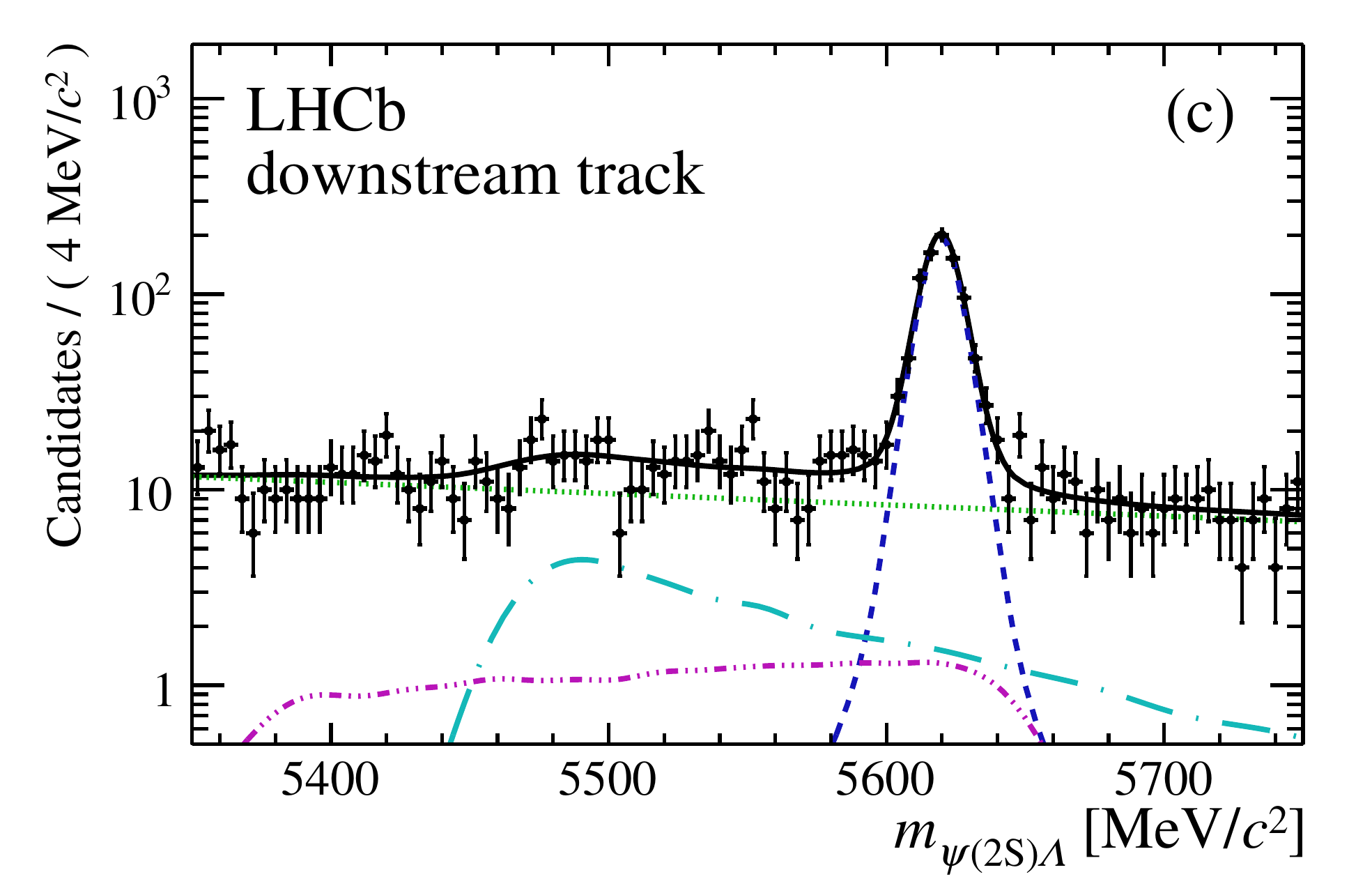}
\includegraphics[width=.5\textwidth]{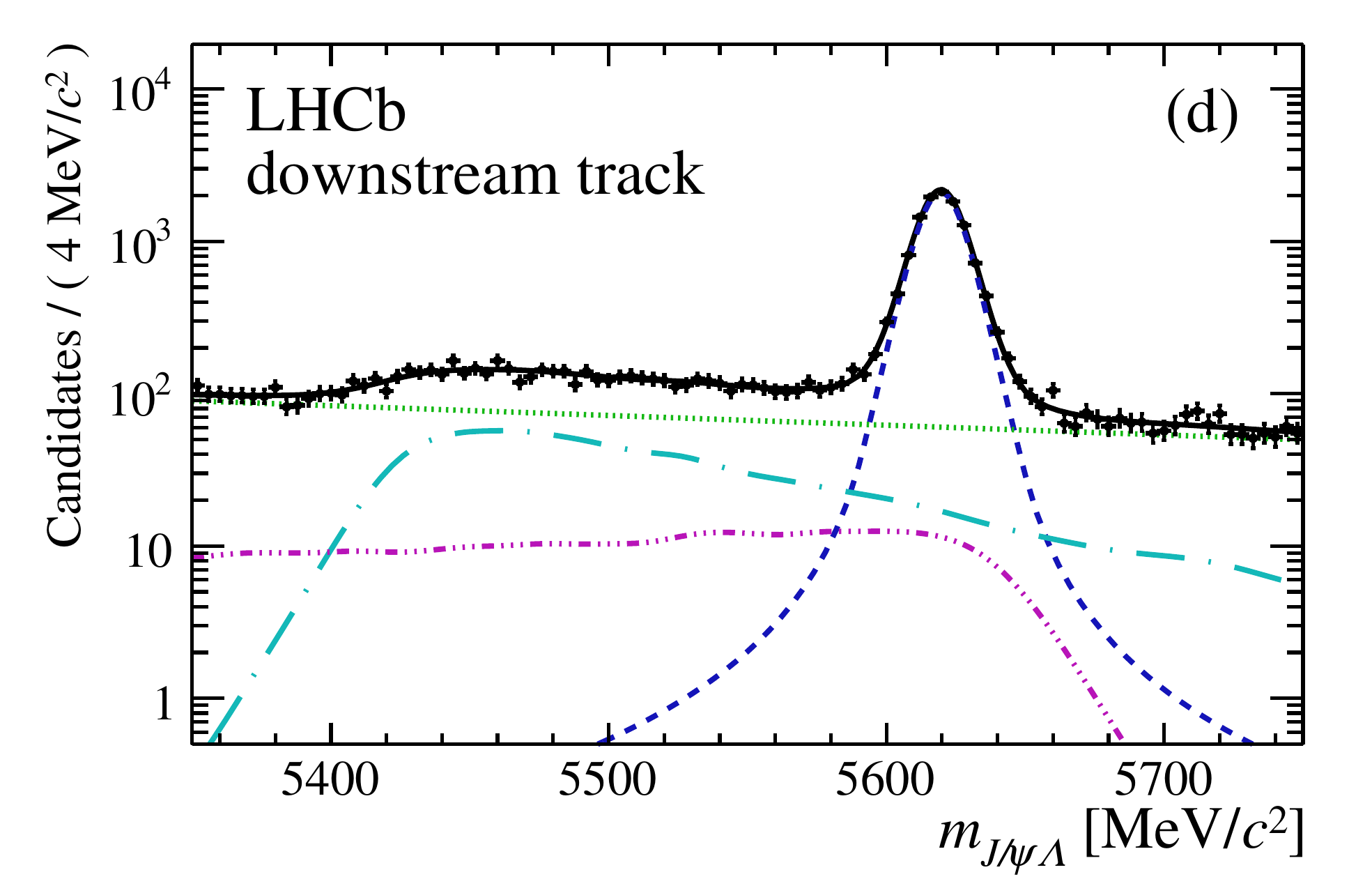}
\caption{Fits to the (unweighted) invariant-mass distributions of long-track candidates for (a) \psichannel and (b) \jpsichannel and for downstream-track candidates for (c) \psichannel and (d) \jpsichannel candidates. The signal (blue, dashed), the combinatorial background (green, dotted), the $\bdccchannel$ background (cyan, long-dash-dotted) and the $\xiccchannel$ background (violet, dash-triple-dotted) are indicated.}
\label{data_fit_unweighted}
\end{figure}

\begin{figure}
\includegraphics[width=.5\textwidth]{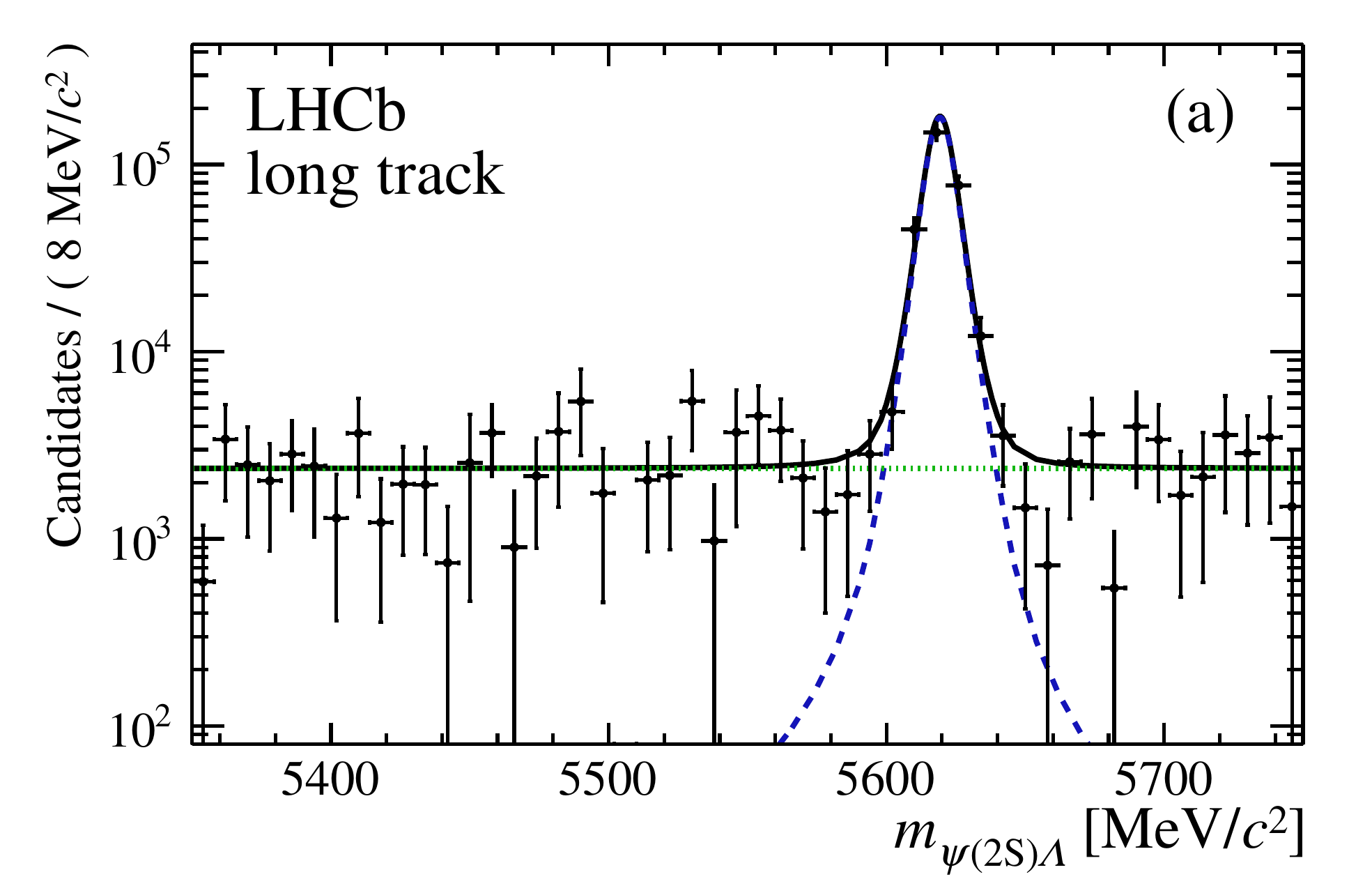}
\includegraphics[width=.5\textwidth]{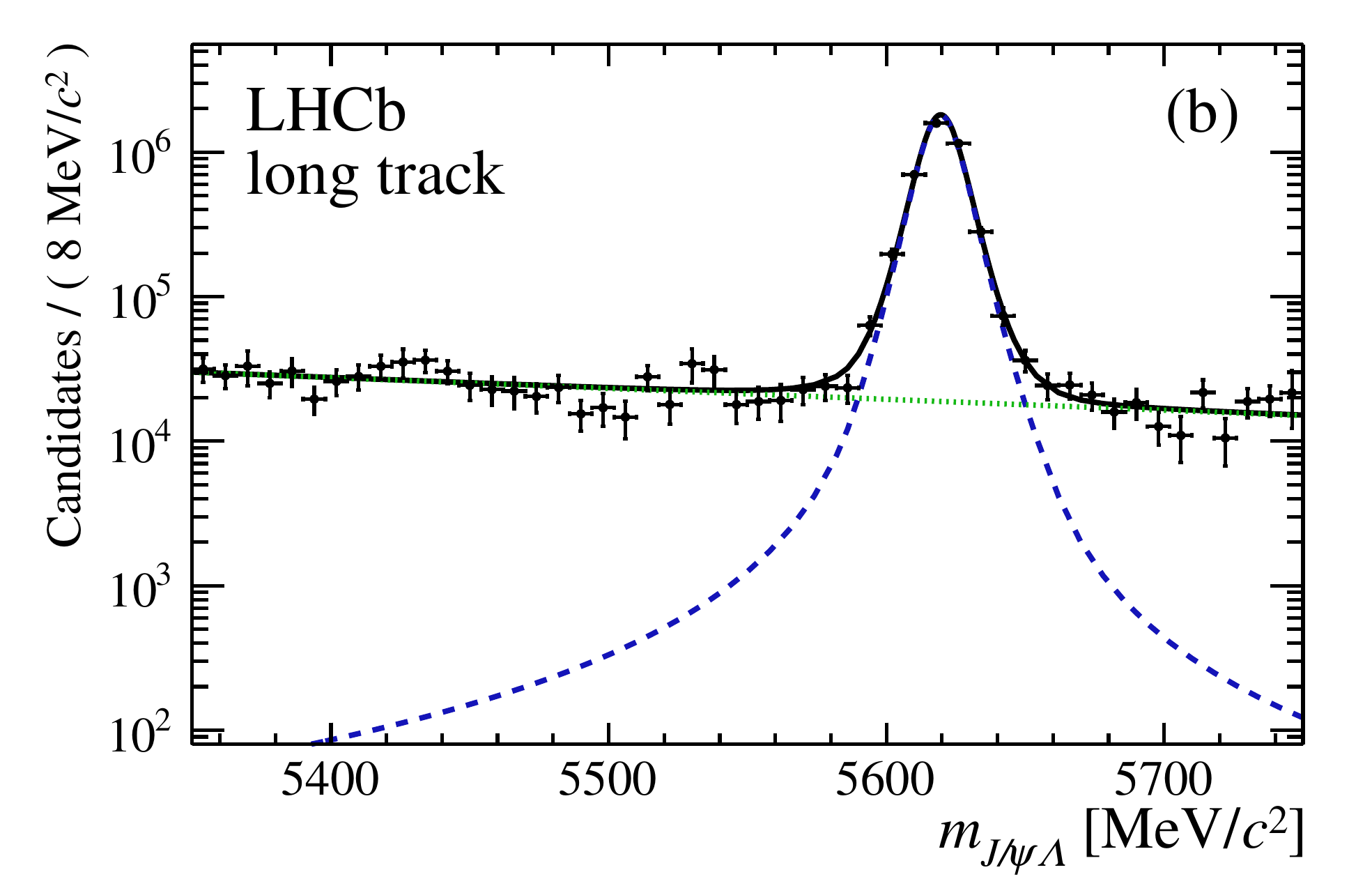}
\includegraphics[width=.5\textwidth]{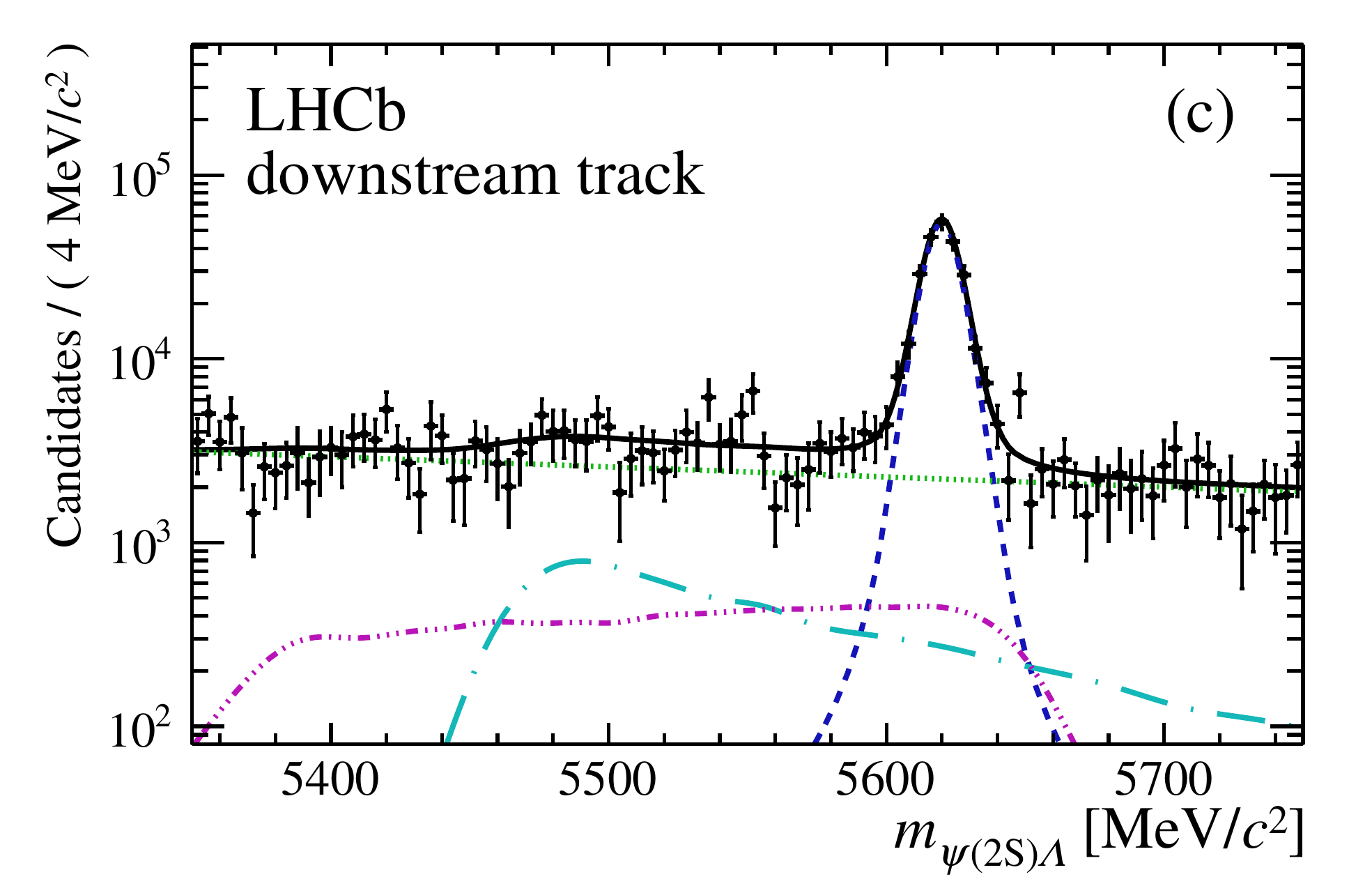}
\includegraphics[width=.5\textwidth]{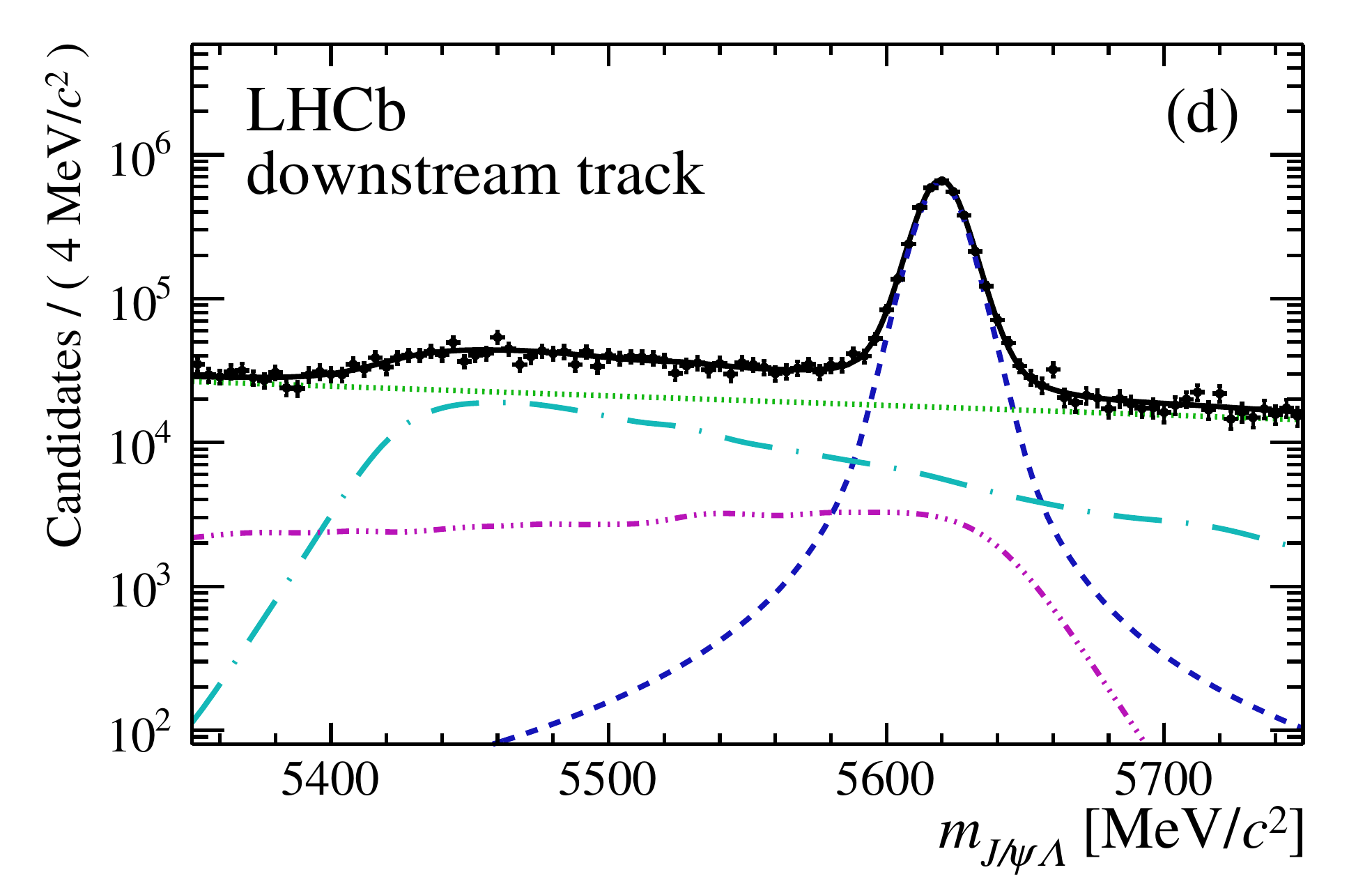}
\caption{Fits to the weighted invariant-mass distributions of long-track candidates for (a) \psichannel and (b) \jpsichannel and for downstream-track candidates for (c) \psichannel and (d) \jpsichannel candidates. The signal (blue, dashed), the combinatorial background (green, dotted), the $\bdccchannel$ background (cyan, long-dash-dotted) and the $\xiccchannel$ background (violet, dash-triple-dotted) are indicated.}
\label{data_fit}
\end{figure}

The signal yield is determined using an extended unbinned maximum likelihood fit to the reconstructed \Lb mass in the range 5350 to 5750\mevcc separately for both decay channels and track types.
The fit model for the reconstructed \Lb mass consists of several components.
The signal is modelled with a double-sided Hypatia function \cite{Santos:2013gra}, where the tail parameters are fixed to values obtained from fits to the simulation.
The combinatorial background is modelled with an exponential function.
A background due to \bdccchannel decays, where the \KS meson decays to two pions and one of the pions is misidentified as a proton, is vetoed in the long-track sample by applying additional particle identification requirements. 
In the downstream-track sample this component is modelled with a kernel-density estimation using a Gaussian kernel \cite{Cranmer:2000du} obtained from simulated \bdccchannel decays.
Another source of background is \xiccchannel decays, where the \Xiresm baryon decays to \L \pim and the pion is not reconstructed.
Contributions from this background source are negligible in the long-track sample due to the sum of the large lifetimes of the \Xires and the \L baryons.
Thus, the \decay{\L}{\proton\pim} decay only happens in less than $2\%$ of the \xiccchannel decays inside the \velo.
In the downstream-track sample this background is modelled with a kernel-density estimation using a Gaussian kernel obtained from simulated \xiccchannel decays.
The number of observed signal events is determined from a fit to unweighted invariant-mass distributions. The resulting fit is shown in \cref{data_fit_unweighted}, separately for long and downstream tracks, and the resulting yields for each data sample are shown in \cref{yields_unweighted}.
In a second fit, the efficiency-corrected yields are obtained assigning to each candidate a weight given by the inverse of the efficiency.
This fit to the two weighted invariant-mass distributions is shown in \cref{data_fit} for each data sample and the resulting efficiency-corrected signal yields for each data sample are reported in \cref{yields}.

\begin{table}[t!]
\center
\caption{Yields from the invariant-mass fits in the range 5350 to 5750\mevcc of (top) $\jpsichannel$ decays and (bottom) $\psichannel$ decays for each component.}
\label{yields_unweighted}
\begin{tabular}{lrrrr}
\toprule
 track type & \jpsichannel & \bdchannel& \xichannel & combinatorial \\ \midrule
downstream& $11\,090 \pm 120$ & $2\,330\pm 210$ &$800 \pm 400$ &$6\,790 \pm 240$ \\
long& $3\,800\pm 60\,\,\,$ & $-$\,\,\,\,\,\,\,\,\,\, & $-$\,\,\,\,\,\,\,\,\,\,&$1\,130 \pm  40\,\,\,$ \\ \midrule
& \psichannel & \bdpsichannel & \xipsichannel & combinatorial \\ \midrule
downstream & $819\pm 33\,\,\,$ & $160\pm 60\,\,\,$ &$60\pm 90\,\,\,$ & $920\pm 60\,\,\,$ \\
long & $317\pm 19\,\,\,$ & $-$\,\,\,\,\,\,\,\,\,\,&$-$\,\,\,\,\,\,\,\,\,\,& $140 \pm 13\,\,\,$ 
 \\\bottomrule
\end{tabular}
\end{table}

\begin{table}[t!]
\center
\caption{Efficiency-corrected yields of $\psichannel$ and $\jpsichannel$ signal decays from the fit to the weighted invariant mass for both track types.}
\label{yields}
\begin{tabular}{lrr}
\toprule
track type & $N_{\psichannel}$ & $N_{\jpsichannel}$ \\ \midrule
downstream & $223\,000\pm 13\,000$ & $3\,320\,000 \pm 50\,000$\\
long & $280\,000\pm 18\,000$ & $3\,980\,000\pm 80\,000$\\\bottomrule
\end{tabular}
\end{table}

\section{Result}
The ratio of branching fractions of \psichannel and \jpsichannel decays is determined separately for long- and downstream-track candidates using
\begin{align}
\label{eq_br}
\frac{\mathcal{B}(\psichannel )}{\mathcal{B}(\jpsichannel )}=\frac{N_{\psichannel}}{N_{\jpsichannel}} \cdot \frac{\mathcal{B}(\jpsi \to \mumu)}{\mathcal{B}(\psitwos \to \mumu)} ,
\end{align}
where $N$ is the number of efficiency-corrected signal candidates, and \mbox{$\mathcal{B}(\jpsi \to \mumu)$} and \mbox{$\mathcal{B}(\psitwos \to \mumu)$} are the known branching fractions of the \Ppsi mesons to two muons \cite{PDG2018}.
Assuming lepton universality, the value for the branching fraction of \psitwos into two electrons, $\mathcal{B}(\psitwos \to \epem)=(0.793\pm 0.017)\%\text{\cite{PDG2018}}$, is used in the calculation due to its lower uncertainty compared to the muon decay.
Using the value for the branching fraction of \jpsi into two muons, $\mathcal{B}(\jpsi \to \mumu) = (5.961\pm 0.033)\%\text{\cite{PDG2018}}$ and the efficiency-corrected signal yields, given in \cref{yields}, the ratios of branching fractions for both track types are calculated to be
\begin{align*}
&\left[\frac{\mathcal{B}(\psichannel )}{\mathcal{B}(\jpsichannel )}\right]_\text{long track}\quad\quad\ \,     = 0.528 \pm 0.036,\\
&\left[\frac{\mathcal{B}(\psichannel )}{\mathcal{B}(\jpsichannel )}\right]_\text{downstream track}= 0.504 \pm 0.029, 
\end{align*}
where the statistical uncertainty only includes the uncertainty on the measured signal yields.
The results for the two classes of tracks are in good agreement and are combined using a weighted average into
\begin{align*}
\frac{\mathcal{B}(\psichannel )}{\mathcal{B}(\jpsichannel )}= 0.513 \pm 0.023.
\end{align*}

\section{Systematic uncertainties}
The sources of systematic uncertainty are summarised in \cref{table_syst}.
The effect of each of these sources on the measured ratio is evaluated independently and is quoted as a relative uncertainty on the measured ratio of branching fractions.
These relative uncertainties are summed in quadrature to obtain the total systematic uncertainty.

\begin{table}
\center
\caption{Relative systematic uncertainties on the ratio of branching fractions.}
\label{table_syst}
\begin{tabular}{lr}
\toprule
 & value\\
\midrule
Simulated dataset size & $1.1\,\%$ \\
Binning choice & $1.6\,\%$\\
Trigger efficiency & $1.2\,\%$\\
Fit model & $1.6\,\%$\\
Simulation correction & $1.3\,\%$\\
$\mathcal{B}(\cquark \cquarkbar\to\lepton\lepton)$ & $2.2\,\%$\\
\midrule
total &  $3.8\,\%$\\
total without $\mathcal{B}(\cquark \cquarkbar\to\lepton\lepton)$&$3.1\,\%$\\ \bottomrule
\end{tabular}
\end{table}

All efficiencies are evaluated from simulated data, therefore the precision is limited by the size of the simulated dataset. 
This effect is determined by varying the binned efficiencies within binomial uncertainties and re-evaluating the efficiency-weighted signal yield.
The result varies by $1.1\,\%$, which is assigned as the systematic uncertainty. 
The effect of the chosen number of bins in both dimensions for the efficiency determination is determined by varying the numbers of bins between five and ten in each dimension independently.
The largest difference compared to the baseline result is a change of $1.6\%$ in the ratio of yields, which is assigned as systematic uncertainty.
To estimate a systematic uncertainty for the trigger efficiency, kinematically similar channels with higher rates, \buchannel and \bupsichannel, are used \cite{LHCb-PUB-2014-039}.
The resulting trigger efficiency on data is compatible with that obtained on simulation, but the systematic uncertainty due to the size of the sample used for this method is $1.2\,\%$.
The effect of using alternative fit models that describe the mass distributions are evaluated using pseudoexperiments.
Candidates are generated using an alternative model and then fitted with the default model.
The $1.6\%$ relative difference between the fitted and generated yield is assigned as systematic uncertainty.
The used correction on the helicity angles in simulation is taken from theory predictions \cite{Gutsche:2013oea}.
An alternative approach is to use the measured distributions from data and this leads to a difference of  $1.3\%$ to the baseline result, which is assigned as systematic uncertainty. 
The effect of neglecting peaking backgrounds for long-track candidates is evaluated by including the \xiccchannel and \bdccchannel components in the long-track sample fits and letting their yields vary freely.
The resulting yields for these components are compatible with zero and the variation of the signal yield is negligible.
Summing these uncertainties in quadrature leads to a systematic uncertainty of $3.1\,\%$.
Another uncertainty arises from the external values for the branching fractions of the charmonium to two muon decays, which is $2.2\,\%$\cite{PDG2018}.

The consistency of the results has been checked by repeating the analysis separately with datasets with different magnet polarities and years of data taking. 
In another crosscheck, the \bdccchannel background is vetoed instead of being included in the fit.
None of these checks shows a significant deviation from the baseline result.

\section{Conclusion}
In summary the ratio of branching fractions is determined to be
\begin{align*}
\frac{\mathcal{B}(\psichannel )}{\mathcal{B}(\jpsichannel )}=  0.513 \pm 0.023 \stat \pm 0.016 \syst \pm 0.011\, (\mathcal{B}),
\end{align*}
where the first uncertainty is statistical, the second is systematic and the third is due to the uncertainty of the used \Ppsi meson branching fractions to two leptons\cite{PDG2018}.
This measurement is compatible within one standard deviation with the measurement from the \atlas collaboration\cite{Aad:2015msa} and has a better precision.
It confirms the discrepancy with the covariant quark model theory predictions\cite{Gutsche:2013oea,Gutsche:2015lea} and sets additional constraints on available models.

\section*{Acknowledgements}
%
%
\noindent We express our gratitude to our colleagues in the CERN
accelerator departments for the excellent performance of the LHC. We
thank the technical and administrative staff at the LHCb
institutes.
We acknowledge support from CERN and from the national agencies:
CAPES, CNPq, FAPERJ and FINEP (Brazil); 
MOST and NSFC (China); 
CNRS/IN2P3 (France); 
BMBF, DFG and MPG (Germany); 
INFN (Italy); 
NWO (Netherlands); 
MNiSW and NCN (Poland); 
MEN/IFA (Romania); 
MSHE (Russia); 
MinECo (Spain); 
SNSF and SER (Switzerland); 
NASU (Ukraine); 
STFC (United Kingdom); 
NSF (USA).
We acknowledge the computing resources that are provided by CERN, IN2P3
(France), KIT and DESY (Germany), INFN (Italy), SURF (Netherlands),
PIC (Spain), GridPP (United Kingdom), RRCKI and Yandex
LLC (Russia), CSCS (Switzerland), IFIN-HH (Romania), CBPF (Brazil),
PL-GRID (Poland) and OSC (USA).
We are indebted to the communities behind the multiple open-source
software packages on which we depend.
Individual groups or members have received support from
AvH Foundation (Germany);
EPLANET, Marie Sk\l{}odowska-Curie Actions and ERC (European Union);
ANR, Labex P2IO and OCEVU, and R\'{e}gion Auvergne-Rh\^{o}ne-Alpes (France);
Key Research Program of Frontier Sciences of CAS, CAS PIFI, and the Thousand Talents Program (China);
RFBR, RSF and Yandex LLC (Russia);
GVA, XuntaGal and GENCAT (Spain);
the Royal Society
and the Leverhulme Trust (United Kingdom);
Laboratory Directed Research and Development program of LANL (USA).



\addcontentsline{toc}{section}{References}
\setboolean{inbibliography}{true}
\bibliographystyle{LHCb}
\bibliography{main,LHCb-PAPER,LHCb-CONF,LHCb-DP,LHCb-TDR}

\newpage
 
\newpage
\centerline
{\large\bf LHCb Collaboration}
\begin
{flushleft}
\small
R.~Aaij$^{42}$,
B.~Adeva$^{41}$,
M.~Adinolfi$^{48}$,
Z.~Ajaltouni$^{5}$,
S.~Akar$^{59}$,
J.~Albrecht$^{10}$,
F.~Alessio$^{42}$,
M.~Alexander$^{53}$,
A.~Alfonso~Albero$^{40}$,
S.~Ali$^{27}$
% ...

{\footnotesize \it

% ...
$ ^{5}$Universit{\'e} Clermont Auvergne, CNRS/IN2P3, LPC, Clermont-Ferrand, France\\
% ...
$ ^{10}$Fakult{\"a}t Physik, Technische Universit{\"a}t Dortmund, Dortmund, Germany\\
% ...
$ ^{27}$Nikhef National Institute for Subatomic Physics, Amsterdam, Netherlands\\
% ...
$ ^{40}$ICCUB, Universitat de Barcelona, Barcelona, Spain\\
$ ^{41}$Instituto Galego de F{\'\i}sica de Altas Enerx{\'\i}as (IGFAE), Universidade de Santiago de Compostela, Santiago de Compostela, Spain\\
$ ^{42}$European Organization for Nuclear Research (CERN), Geneva, Switzerland\\
% ...
$ ^{48}$H.H. Wills Physics Laboratory, University of Bristol, Bristol, United Kingdom\\
% ...
$ ^{53}$School of Physics and Astronomy, University of Glasgow, Glasgow, United Kingdom\\
% ...
$ ^{59}$University of Cincinnati, Cincinnati, OH, United States\\
% ...
}
\end{flushleft}

\end{document}